\makeatletter
\@ifundefined{@parse@version@dash}{%
\def\@parse@version#1{\@parse@version@0#1}
\def\@parse@version@#1/#2/#3#4#5\@nil{%
\@parse@version@dash#1-#2-#3#4\@nil}
\def\@parse@version@dash#1-#2-#3#4#5\@nil{%
  \if\relax#2\relax\else#1\fi#2#3#4 }
}{}
\makeatother

\documentclass[prl,twocolumn,superscriptaddress,showpacs,floatfix,longbibliography]{revtex4-1}
\usepackage{mathrsfs,braket}
\usepackage{amssymb, amsbsy, amsmath, latexsym, dsfont, array, layout, graphicx,mathrsfs,color,ulem,bm,soul}
\usepackage[colorlinks=true,citecolor=blue,urlcolor=blue,linkcolor=blue]{hyperref}

\begin{document}
\title{Engineering topological chiral transport in a flat-band lattice of ultracold atoms}

\author{Hang Li}
\thanks{These authors contributed equally to this work}
\author{Qian Liang}
\thanks{These authors contributed equally to this work}
\affiliation{%
Zhejiang Province Key Laboratory of Quantum Technology and Device, School of Physics, and State Key Laboratory for Extreme Photonics and Instrumentation, Zhejiang University, Hangzhou 310027, China
}%
\author{Zhaoli Dong}
\author{Hongru Wang}
\affiliation{%
Zhejiang Province Key Laboratory of Quantum Technology and Device, School of Physics, and State Key Laboratory for Extreme Photonics and Instrumentation, Zhejiang University, Hangzhou 310027, China
}%
\author{Wei Yi}
\affiliation{CAS Key Laboratory of Quantum Information, University of Science and Technology of China, Hefei 230026, China}
\affiliation{CAS Center For Excellence in Quantum Information and Quantum Physics, Hefei 230026, China}
\author{Jian-Song Pan}
\email{panjsong@scu.edu.cn}
\affiliation{College of Physics and Key Laboratory of High Energy Density Physics and Technology of Ministry of Education, Sichuan University, Chengdu 610065, China}
\author{Bo Yan}
\email{yanbohang@zju.edu.cn}
\affiliation{%
Zhejiang Province Key Laboratory of Quantum Technology and Device, School of Physics, and State Key Laboratory for Extreme Photonics and Instrumentation, Zhejiang University, Hangzhou 310027, China
}%
\affiliation{%
College of Optical Science and Engineering, Zhejiang University, Hangzhou 310027, China
}%

\date{\today}

\begin{abstract}

The manipulation of particle transport in synthetic quantum matter is an active research frontier for its theoretical importance and potential applications.
Here we experimentally demonstrate an engineered topological transport in a synthetic flat-band lattice of ultracold $^{87}$Rb atoms.
We implement a quasi-one-dimensional rhombic chain with staggered flux in the momentum space of the atomic condensate and observe biased local oscillations that originate from the interplay of the staggered flux and flat-band localization under the mechanism of Aharonov-Bohm caging.
Based on these features, we design and experimentally confirm a state-dependent chiral transport under the periodic modulation of the synthetic flux. We show that the phenomenon is topologically protected by the winding of the Floquet Bloch bands of a coarse-grained effective Hamiltonian.
The observed chiral transport offers a strategy for efficient quantum device design where topological robustness is ensured by fast Floquet driving and flat-band localization.

\end{abstract}

\maketitle

{\bf Introduction.}
Quantum control is an essential ingredient of modern quantum science and technology. The quest of manipulating localization and transport dynamics in particular has significantly enriched our understanding of quantum matter~\cite{Anderson,2010_RMPTI1,2011_RMPTI2}, with ample implications for quantum device design~\cite{2012_Supercon,2012_ions,2012_qugas,Goldman2016,2019_RMP1,2020_Rydberg,2019_RMP2,2013_NVreview}.
With flexible tunability in the lattice geometry, site-resolved hopping patterns, disorder, and synthetic flux, cold atoms in a synthetic momentum lattice have developed into an important platform for simulating and exploring transport dynamics in a wealth of physical contexts~\cite{Schafer2020,Cooper2019}.
Over the past decade, momentum-lattice engineering has helped to reveal rich quantum dynamic phenomena regarding topological matter~\cite{Meier2016a,2019_npj}, non-Hermitian physics~\cite{Gou2020,Liang2022}, correlated dynamics in frustrated geometry~\cite{Ma2023nc}, as well as localization and mobility edges~\cite{PhysRevX.8.031045,PhysRevLett.126.040603,PhysRevLett.129.103401}.
A prime example here is the experimental demonstration of the inverse Anderson localization~\cite{Inverse_prl,Inverse_ol,hang2022prl,zhang2022prb,zhang2023prl}, where disorder is shown to induce delocalization and particle transfer in the presence of flux-induced flat bands under the Aharonov-Bohm (AB) caging. A natural question is whether the interplay of synthetic flux and flat-band localization can be harnessed for the switching and control of the transport dynamics.

In this work, we experimentally demonstrate topological chiral transport in a synthetic rhombic chain in the momentum space of ultracold $^{87}$Rb atoms.
Adopting the state-of-the-art momentum-lattice engineering technique, we are able to switch and tune the synthetic flux threaded through each plaquette on demand and probe the dynamics of the condensate along the synthetic lattice.
We observe local breathing modes and biased oscillations in a binary staggered-flux configuration, both phenomena deriving from the interplay of the synthetic flux and the flat-band localization under the mechanism of AB caging.
We then engineer a state-dependent chiral transport by introducing Floquet modulations to the flux. While the micromotion of the Floquet dynamics involves the occupation of multiple sublattice sites and is complicated in general, we show that the chiral transport is quantized, and characterized by the winding of the Floquet Bloch bands of a coarse-grained effective Hamiltonian. While the topological robustness and fast time scale of the quantized transport can be useful for quantum device design, our setup further paves the way for the exploration of the rich dynamics under the interplay of flat-band localization, disorder, Floquet driving, and long-range interactions typical of the momentum lattice.

\begin{figure}[!t]
	\centering
	\includegraphics[width=1\linewidth]{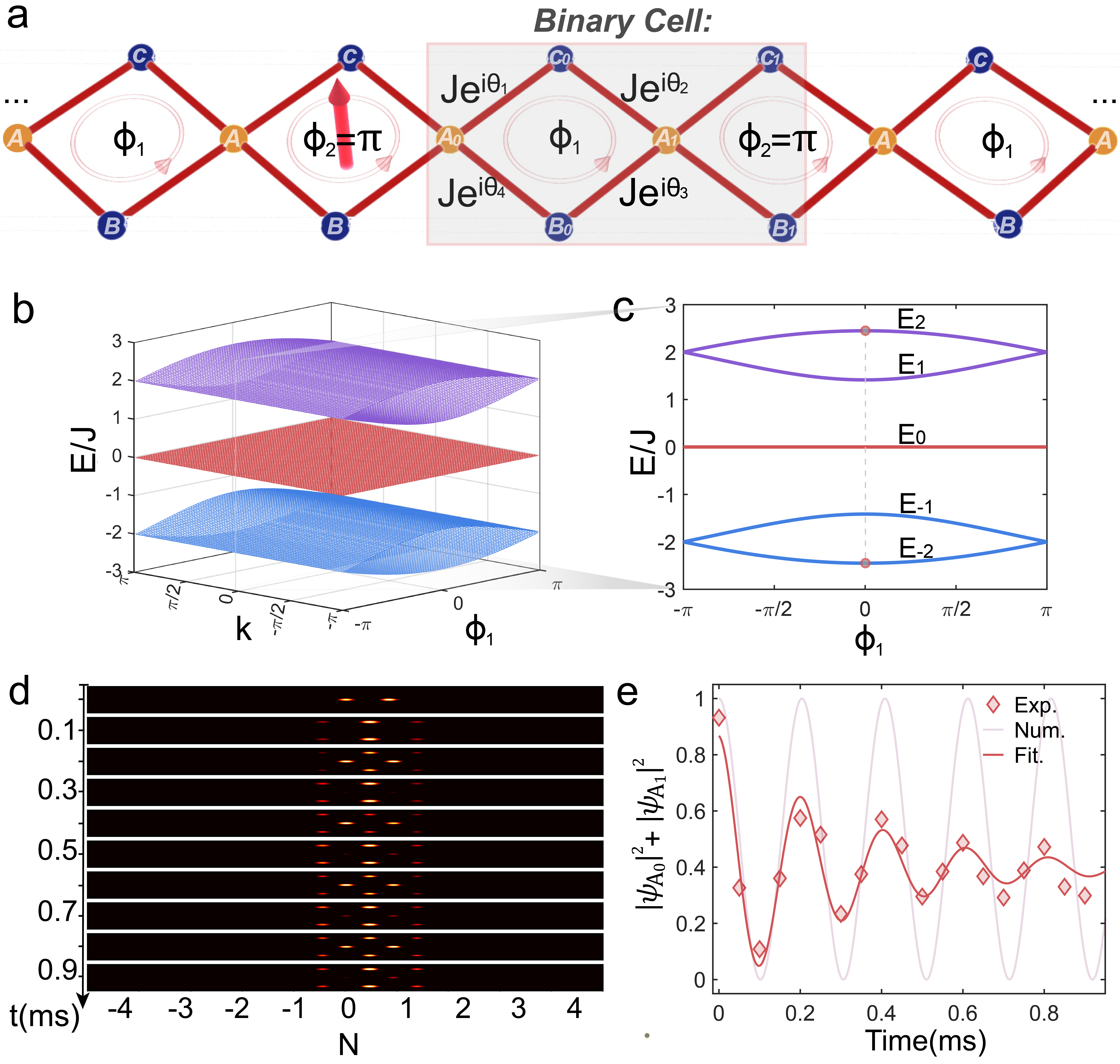}
\caption{\textbf{Schematic of the rhombic flux-staggered chain and the breathing mode.} \textbf{a.} Schematics of a rhombic flux-staggered chain with flexibly programmable artificial gauge fields.  The subfigure shows the binary flux-staggered ladder configuration with a static synthetic flux setting. The complex hopping coefficients between the nearest-neighbor sites are denoted as $Je^{i\theta_{m}} (m=1,2,3,4)$. \textbf{b.} The energy dispersion $E(k)$ of the binary-flux lattice chain with $\phi_{2}=\pi$. All the bands display complete flat-band features no matter what $\phi_1$ is. \textbf{c.} The flat-band splitting with the variation of $\phi_{1}$. When $\phi_{1} \neq \pm \pi$, the binary-flux lattice has six flat bands, labeled as $E_{\pm2}, E_{\pm1}, E_0$ respectively. The central band with $E_0$ is two-fold degenerate.  \textbf{d.} The evolution dynamics for an initial state $\left|\psi_{\text{ini}} \right>=\frac{1}{\sqrt{2}}(\hat{a}_{n}^{\dagger}+\hat{a}_{n+1}^{\dagger})\left| 0\right\rangle$ with the binary-flux parameter \{$\phi_{1} = 0, \phi_{2} = \pi$\}. \textbf{e.} Measuring a damped oscillation of the total population of sites $A_{0}$ and $A_{1}$ around $0.9$ ms. The red solid lines represent the fitting of experimental data, and the pale red line is the numerical results calculated using $H_\text{eff}$ in Eq.~(\ref{Heff}). }
	\label{scheme}
\end{figure}

{\bf Implementing the staggered-flux lattice.}
To study the interplay of flat-band and synthetic flux, we consider a quasi-1D rhombic chain with individually controllable synthetic flux threaded through each plaquette, as shown in Fig.~1(a). Under the tight-binding approximation~\cite{2015_pra,2016_pra}, the system can be described by the effective Hamiltonian
\begin{equation}\label{Heff}
\begin{aligned}
&H_\text{eff}=\sum_{n}[-J(e^{i\theta_{n-1,3}}\hat{a}_{n}^{\dagger}\hat{b}_{n-1}+e^{-i\theta_{n,4}}\hat{a}_{n}^{\dagger}\hat{b}_{n}+\\& \ \ \ e^{-i\theta_{n-1,2}}\hat{a}_{n}^{\dagger}\hat{c}_{n-1}+e^{i\theta_{n,1}}\hat{a}_{n}^{\dagger}\hat{c}_{n}+\text{H.c.})],
\end{aligned}
\end{equation}
where $J$ is the hopping amplitude between adjacent sites, $\hat{a}_{n}^{\dagger}$, $\hat{b}_{n}^{\dagger}$ and $\hat{c}_{n}^{\dagger}$ ($\hat{a}_{n}$, $\hat{b}_{n}$ and $\hat{c}_{n}$) are the creation (annihilation) operators for atoms on the sublattice sites $A_{n}$, $B_{n}$, and $C_{n}$, respectively.
Since the synthetic flux of the $n$-th plaquette $\phi_n$ is given by $\phi_{n}(t)=\sum_{m=i}^{4} \theta_{n,m}(t)$, we henceforth adopt the gauge convention: $\theta_{n,2}=\phi(t)$, and $\theta_{n,m}=0$ otherwise.

Throughout the work, we focus on a binary-flux-ladder (BFL) configuration with recurring
$\phi_{1}$ and $\phi_{2}$, as illustrated in Fig.~1(a). Each unit cell contains $6$ sublattice sites, respectively labeled as $\{A_{n}, B_{n}, C_{n}, A_{n+1}, B_{n+1}, C_{n+1}\}$. Here $\{A_{n}, B_{n}, C_{n}, A_{n+1}\}$ encircle the $n$-th plaquette, consistent with Eq.~(\ref{Heff}).
The band dispersions are
$E_{0}=0$, $E_{\pm1}=\pm J\sqrt{4- \sqrt{2\left ( 2+\cos(\phi _{1})+\cos(\phi _{2})+F(\phi _{1},\phi _{2}) \right )}}$, and $E_{\pm2}=\pm J\sqrt{4+ \sqrt{2\left ( 2+\cos(\phi _{1})+\cos(\phi _{2})+F(\phi _{1},\phi _{2}) \right )}}$. Here the flux-dependent factor $F(\phi _{1},\phi _{2})=4\cos(2k)\cos(\phi_1/2)\cos(\phi_2/2)$, $k \; (-\pi \leq k< \pi )$ is the Bloch wave number. For convenience, we have set the lattice constant to unity.
Notably, when $\phi_{2}=\pi$, $F(\phi _{1},\phi _{2})=0$ for arbitrary $\phi_1$. All the bands are then flat regardless of $\phi_1$ [see Fig.~1(b)]. This gives rise to intricate localized states and opens up room for engineering the transport dynamics, as we show below.

Experimentally, we implement the rhombic chain in Fig.~1(a) along a Raman-coupled momentum lattice in a $^{87}\textrm{Rb}$ Bose-Einstein condensate (BEC), which contains $\sim 2\times10^{5}$ atoms~\cite{2018_JOSAB,2021_npj,Dong2023}. As illustrated in Figs.~1(a) and 1(b), sublattice sites $\{A_n, B_n, C_n\}$ are encoded in the momentum states of atomic hyperfine levels~\cite{supp}: $A_n$ and $B_n$ correspond to different momentum states of $\left| F=1,m_{F}=0 \right\rangle$, and $C_n$ is encoded in those of $\left| F=2,m_{F}=0 \right\rangle$, respectively.
Hopping between adjacent momentum-lattice sites is implemented by two-photon Raman or Bragg processes~\cite{hang2022prl,supp}, with a fixed hopping amplitude $J= 0.95(5)$ kHz. The synthetic flux $\phi_n$ is controlled by locking the relative phases of the coupling lasers~\cite{hang2022prl},
and all $\phi_n$ can be individually controlled.

{\bf Breathing mode and biased oscillation.}
An outstanding dynamic signature of the flat-band configuration is the localized breathing mode under the AB-caging mechanism~\cite{ABcage_prl, ABcage_nc,ABcage_prl2022,hang2022prl}. The conventional AB caging occurs for $\phi_1=\phi_2=\pi$, but persists under the BFL configurations with only $\phi_2=\pi$. We experimentally confirm this by focusing on the case with $\phi_{1}=0$ and $\phi_{2}=\pi$, which has the largest gap between the low- and high-lying bands $E_{\pm 2}$.
The corresponding localized eigenstates of the two flat bands are $\left| \psi_{n,\pm2}^{\text{}} \right\rangle=\frac{1}{2}[(\hat{a}^{\dagger}_{n}+\hat{a}^{\dagger}_{n+1})\pm \frac{\sqrt{6}}{6}(2\hat{b}^{\dagger}_{n}+2\hat{c}^{\dagger}_{n} -\hat{b}^{\dagger}_{n-1}+ \hat{c}^{\dagger}_{n-1}+\hat{b}^{\dagger}_{n+1}+\hat{c}^{\dagger}_{n+1})]\left| 0 \right\rangle$.
As such, an initial state $\left|\psi_{\text{ini}} \right\rangle=\frac{1}{\sqrt{2}}(\hat{a}_{n}^{\dagger}+\hat{a}_{n+1}^{\dagger})\left| 0\right\rangle$ can be expressed as an equal-weight superposition of $\left| \psi_{n,\pm2}^{\text{}} \right\rangle$, and the subsequent breathing mode should acquire a frequency proportional to the energy gap between the bands.

In the experiment, we prepare the state $\left|\psi_{\text{ini}} \right\rangle$ by splitting the initial BEC wave packet on-site $B_0$ (with the plaquette label $n=0$), and show the detected density dynamics in Fig.~1(d).
While the ensuing dynamics is localized and centered around the initial site, the breathing-mode character is clearly visible in Fig.~1(e).
The observed breathing mode features a frequency $\omega=$4.9(2) kHz, consistent with
the numerically calculated energy gap $\omega=4.9J$ in Fig.~1(c).
Due to the decoherence of the Raman-coupling processes, the overall dynamics is damped, with a fitted decay lifetime of 0.40(5)ms, as illustrated in Fig.~1(e).

\begin{figure}[t]
	\centering
	\includegraphics[width=1.0\linewidth]{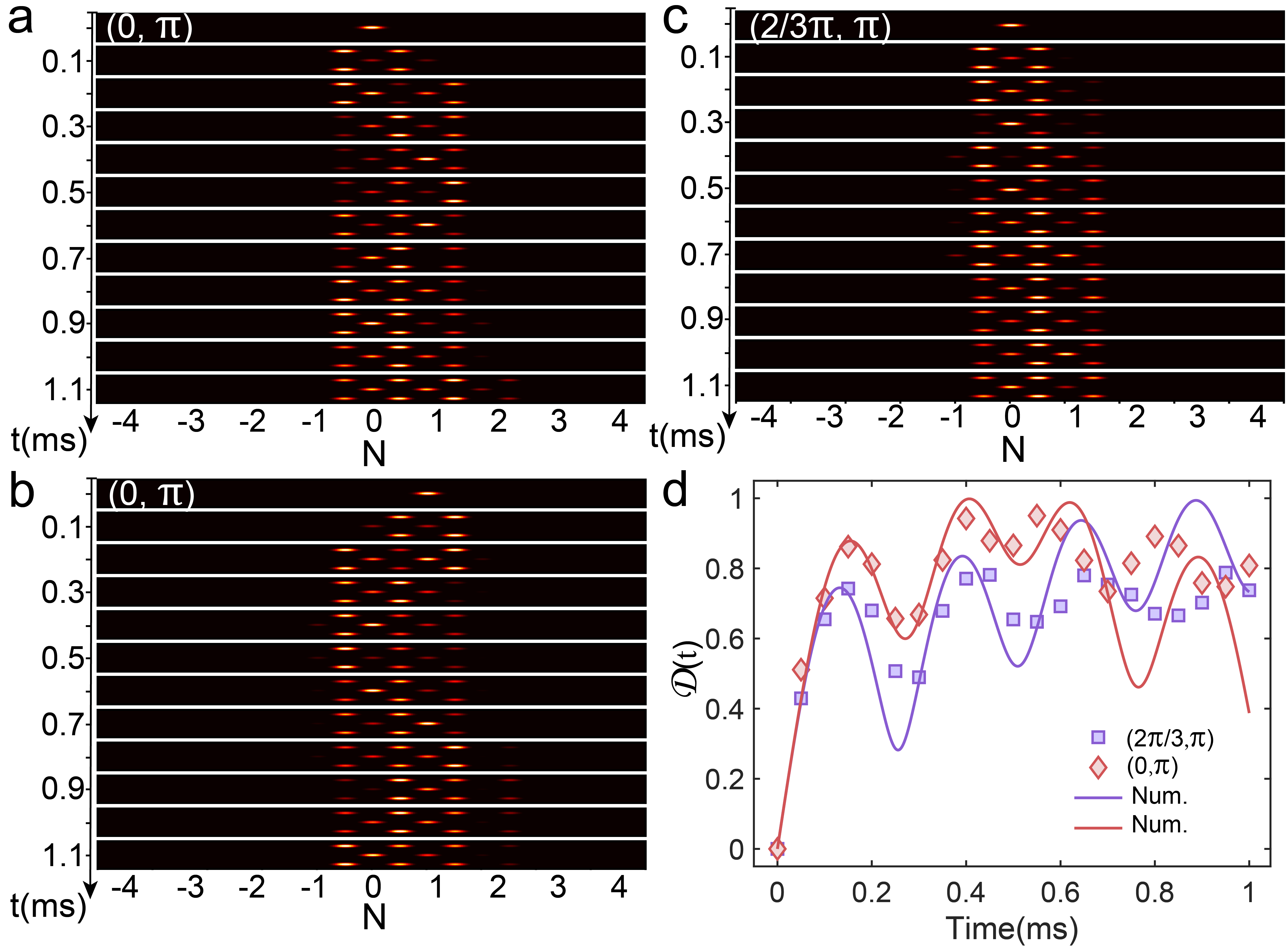}
	\caption{\textbf{Biased oscillations in the binary flux-staggered lattice.}  \textbf{a.} The evolution dynamics under \{$\phi_{1} = 0, \phi_2=\pi$\}, with the initial state $\psi_\text{ini}=\hat{a}_0^{\dagger}|0\rangle$. \textbf{b.} The evolution dynamics under \{$\phi_{1} = 0, , \phi_2=\pi$\} with $\psi_\text{ini}=\hat{a}_1^{\dagger}|0\rangle$. \textbf{c.} The evolution dynamics under \{$\phi_{1} = 2\pi/3, , \phi_2=\pi$\} with $\psi_\text{ini}=\hat{a}_0^{\dagger}|0\rangle$. \textbf{d.} The extracted $\mathcal{D}(t)$ with $\psi_\text{ini}=\hat{a}_0^{\dagger}|0\rangle$ in different BFL parameters setting. The dots are experimental data, and solid lines are numerical simulations using $H_\text{eff}$.
	}
        \label{figure2}
\end{figure}

For a more general choice of $\phi_1$ and local initial state, the dynamics of the condensate center of mass becomes oscillatory but is biased in direction depending on the initial state.
For instance, Figs.~2(a) and 2(b) display the density dynamics under $\{\phi_{1} = 0,\phi_2=\pi\}$ with different initial-site excitations.
While the dynamics is largely localized within the initial unit cell under the flat-band condition (enforced by $\phi_2=\pi$), atoms initialized on-site $A_n$ ($A_{n+1}$) move toward the right (left) at the beginning of the time evolution, and the
overall oscillation is accordingly biased in direction.
Further, by varying $\phi_1$, we can adjust the frequency of the local oscillation, as shown
in Fig.~2(c). This becomes clearer in Fig.~2(d), where we plot the second moment of the position operator, defined through $\mathcal{D}^{2}(t)= \sum_{n}(n-n_0)^{2}(\left| a_{n}\right|^{2}+\left| b_{n}\right|^{2}+\left| c_{n}\right|^{2})$, where $n_0$ is the position of initial injection state, $a_{n}$, $b_{n}$ and $c_{n}$ are the expectation values of $\hat{a}_{n}$, $\hat{b}_{n}$ and $\hat{c}_{n}$, respectively. While $\mathcal{D}(t)$ is often used to indicate the transport velocity and classify quantum transport ~\cite{Inverse_ol,hang2022prl}, we observe that the frequency and range of $\mathcal{D}(t)$ are tunable through $\phi_1$.

\begin{figure*}[!t]
\centering
\includegraphics[width=1.0\textwidth]{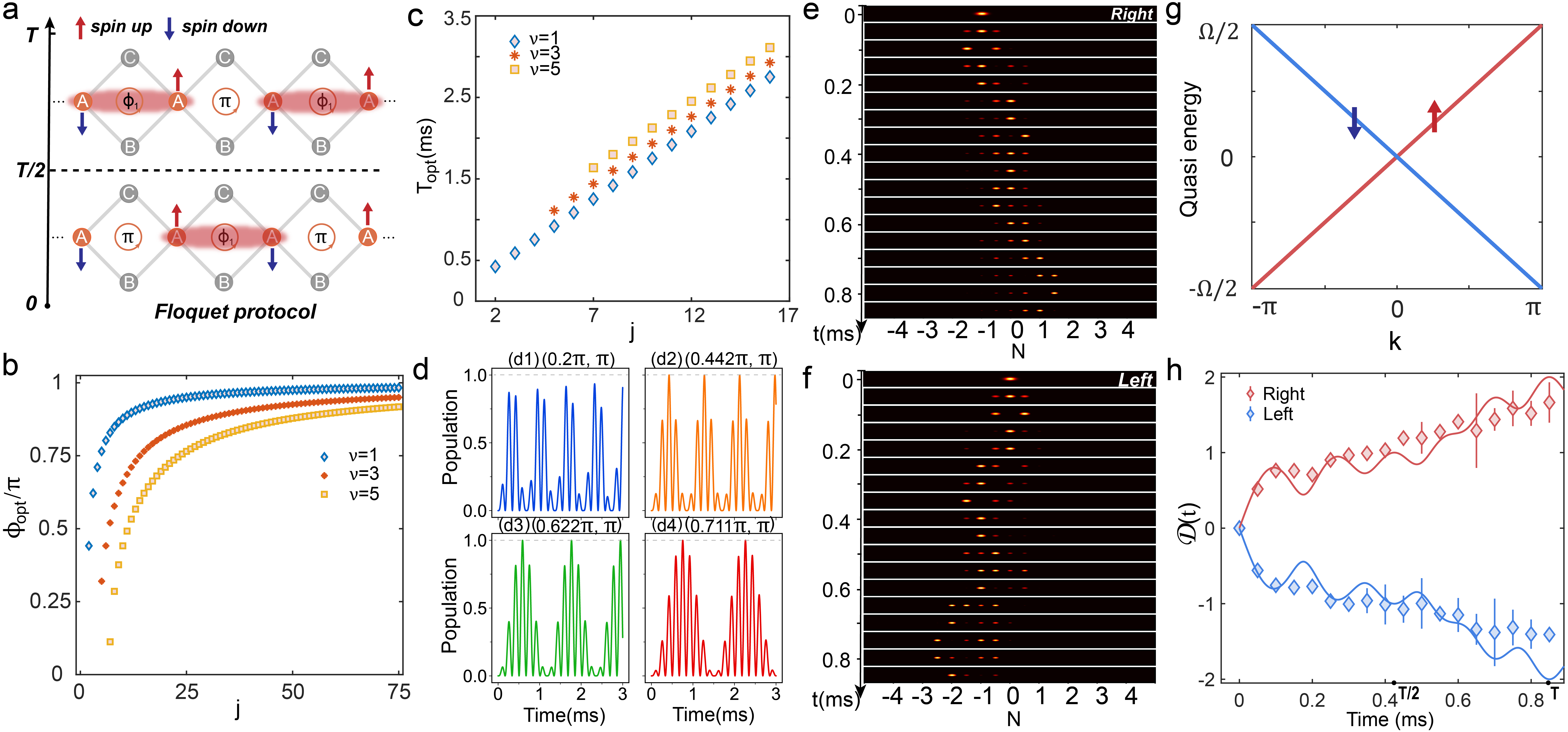}
\caption{\textbf{Floquet channel and topological chiral transport.} \textbf{a.} The Floquet protocol by exchanging the adjacent synthetic flux at the spectroscopic moments. In the first half T/2 period, the system can be described by the Hamiltonian $H_1$; In the second half T/2 period, the system is switched to the Hamiltonian $H_2$. \textbf{b.} The theoretical results of optimal flux setting with perfect population transfer. \textbf{c.} The calculated Floquet period time $T_\text{opt}$ with perfect transfer corresponding to each $\phi_{\text{opt}}$. \textbf{d.} The transfer dynamics within the spinor pair with different $\phi_{1}$ settings. The subfigure (d1) shows the population transfer under the \{$\phi_{1}=0.2\pi, \phi_{2}=\pi $\} BFL configuration. The last three subfigures (d2-d4) are corresponding to different optimal $\phi_{1}=\phi_{\text{opt}}$ settings with $j$=1, $\nu=$2, 3, 4, respectively ($\phi_{\text{opt},11}$ is an imaginary number and thus is irrelevant).  \textbf{e.} and \textbf{f.} The experimental results of the Floquet channel with right and left chiral transport, respectively. For the reason that we mainly observe the chiral current represented by the spinor pair sites $A_{n}$, it is unnecessary to distinguish the micromotions of atomic density in the $B_{n}$ and $C_{n}$ sites. So, we have combined the display of $B_{n}$ and $C_{n}$ sites by indicating their total density information. \textbf{g.} The quasi-energy spectrum of the Floquet channel protocol. \textbf{h.} The extracted $\mathcal{D}(t)$ curve of experimental result for the chiral transports of the Floquet channels. The solid lines show the numerical simulations. The error bars present the standard deviation of measurements.
}
\label{figure3}
\end{figure*}

{\bf Topological chiral transport.}
The local dynamics observed above can be harnessed for quantized chiral transport through the Floquet engineering.
We show that a topological chiral transport can be achieved by constructing a helical Floquet channel ~\cite{Floquetch} which leads to a perfect spin-momentum locking through the winding of the Floquet Bloch bands.

Specifically, assuming the wavefunction is initialized on a local sublattice site $A_n$, under the interplay of flat-band localization and staggered flux, a perfect population transfer between sites
$A_n$ and $A_{n+1}$ can be achieved under appropriate flux parameters and at a discrete set of times.
Analytically, we find that, following an evolution time $t$, the wavefunction on
$A_n$ and $A_{n+1}$ are respectively given as $[\cos(\varepsilon _{1}t)+\cos(\varepsilon _{2}t)]/2$ and $[\cos(\varepsilon _{1}t)-\cos(\varepsilon _{2}t)]/2$, where $\varepsilon _{1}$ and $\varepsilon _{2}$ are the eigen energies of real-space Hamiltonian~\cite{supp}. It follows that a perfect population transfer occurs at $T_{\text{opt},j\nu}$, under the conditions $\varepsilon _{1}T_{\text{opt},j\nu}=(j+\nu)\pi$ and $\varepsilon _{2}T_{\text{opt},j\nu}=j\pi$, where $j$ is a positive integer and $\nu$ is a positive odd integer.
In terms of the flux parameters, the conditions for the perfect population transfer are given by
\{$\phi_{1}=\phi_{\text{opt},j\nu}, \phi_{2}=\pi$\}, with
\begin{equation}\label{eq:phi_m}
\phi_{\text{opt},j\nu}=2\arccos\left[\frac{2\nu(2j+\nu)}{j^{2}+(j+\nu)^{2}}\right], \,\,\,\text{and}\,\,\, j\geq \frac{1+\sqrt{3}}{2}\nu.
\end{equation}
In Fig.~3(b), we show the discrete values of $\phi_{\text{opt},j\nu}$ that satisfy the conditions above.
On the other hand, we also have
\begin{equation}\label{eq:T_m}
T_{\text{opt},j\nu}=\frac{j\pi}{J\sqrt{2(2-\cos(\phi_{opt,j\nu}/2))}},
\end{equation}
which gives the time of the perfect population transfer, as shown in Fig.~3(c).

Note that the above conditions are highly non-trivial, as the full dynamics generically involve the occupation of all the sublattice sites including $B_n$ and $C_n$. Qualitatively, the perfect population transfer derives from the interference between different flat-band components of the initial state.

Based on the flux-dependent local dynamics, we propose to implement a spin-dependent chiral transport as follows.
First, we initialize the condensate on-site $A_n$ to the left of the plaquette with flux $\phi_{\text{opt},j\nu}$. We then let the condensate evolve along the momentum lattice, switching the synthetic fluxes $\phi_1$ and $\phi_2$ within each unit cell at integer multiples of $T_{\text{opt},j\nu}$.
Taking $\nu=1$ and $j=2$ for instance, for a wavefunction initialized on-site $A_n$ under the synthetic-flux configuration \{$\phi_{1} =\phi_{\text{opt},21}\approx 0.442\pi, \phi_{2} = \pi$\}, it propagates to the right and becomes fully localized on-site $A_{n+1}$ at
$t=T_{\text{opt},21}\approx0.425$ ms when the coupling $J=1.5$ kHz. We then swap $\phi_1$ and $\phi_2$ so that the wavefunction continues to propagate toward the right, becoming fully localized on the central site of the next plaquette (site $A_{n+2}$) at $t=2T_{\text{opt},21}$. By repeating this procedure, a persistent chiral current is realized, which is quantized at discrete time steps.
By contrast, the transport is not quantized when, for instance, \{$\phi_{1} = 0.2\pi, \phi_{2} = \pi$\}, since the oscillatory dynamics become rather complicated, as illustrated in Fig.~3(d).

For the experimental confirmation of the chiral dynamics, we excite both spin-up and spin-down components at $t=0$, and
observe the right- and left-going chiral transport in Fig.~3(e) and 3(f), respectively.
The quantized chiral transport can be understood by neglecting the complicated micromotion, and
focusing on coarse-grained Floquet dynamics at discrete time steps $t=2mT_{opt,j\nu}$ ($m=0,1,2...$).
The subsequent dynamics involve only the occupation of the central two sites $A_n$ and $A_{n+1}$ in each unit cell so that
we can map them to pseudospins, with the spinor field operator $\psi _{n}=(\psi _{n\uparrow},\psi_{n\downarrow} )$.
The stroboscopic Floquet dynamics in the two-dimensional spinor subspace can be described by the Floquet operator
$U=e^{-iH_{2} T/2}e^{-iH_1T/2}$, where $H_{1}=\frac{\pi }{T}\sum_{n}\psi _{n}^{\dagger }\sigma _{x}\psi _{n}$, $H_{2}=-\frac{\pi }{T}\sum_{n}\psi _{n}^{\dagger }S^{+}\psi _{n}$, and $S^{+}=\frac{1}{2}(\sigma _{x}+i\sigma _{y})$. Here $T=2T_{\text{opt},j\nu}$, and $H_{1}$ and $H_{2}$ respectively describe the spin-flip process under the switching of flux and the transport process during each half period.
In the quasi-momentum space, we have $U_{k}(T,0)=e^{-ik\sigma _{z}}$~\cite{Floquetch}, and it follows that the Floquet Bloch Hamiltonian reads~\cite{supp}

\begin{equation}\label{eq4}
\begin{aligned}
H_{k}^{F}=\frac{i}{T}log[U_{k}(T,0)]=\frac{1}{T}k\sigma _{z}.
\end{aligned}
\end{equation}
The Hamiltonian $H_{k}^{F}$ has the intrinsic feature of spin-momentum locking in the real space, which underlies the spin-dependent chiral transport.
Indeed, this unique feature can be revealed by the Floquet band structure, shown in Fig. 3(g),
where the quasienergy spectrum is gapless, featuring decoupled linear dispersions for the two spin species. We note that these helical Floquet channels exist at discrete time steps when the interim occupations of sites $B_n$ and $C_n$ are neglected. This is different from previous studies where simpler engineered SSH lattice models were used~\cite{Xiao2020,PhysRevLett.129.040402}.

In Fig.~3(h), we show the measured chiral transport using the quantity $\mathcal{D}(t)$. While different spin species flow in different directions, quantized
chiral transport is achieved at discrete times $mT/2$ ($m=0,1,2,...$), consistent with the description of the stroboscopic Floquet dynamics.
The quantized transport is topologically protected by the Floquet winding numbers~\cite{Floquetch,PhysRevB.82.235114}
\begin{equation}\label{eq:winding_number}
\nu_{\sigma}=\frac{1}{2\pi i}\oint_{BZ}dk\text{Tr}[U_{k}^{\sigma}\partial_{k}U_{k}^{\sigma\dagger}],
\end{equation}
where $U_{k}^{\uparrow,\downarrow}=e^{\pm ik}$ are the irreducible blocks of $U_{k}$.
The Floquet winding numbers reflect the state-dependent winding of quasi-energy bands as the quasi-momentum $k$ traverses the Brillouin zone.
Consistent with Fig.~3(g), we find $\nu_{\uparrow}=1$ ($\nu_{\downarrow}=-1$), indicating quantized transport to the right (left). The Floquet chiral dynamics is hence topological.

{\bf Conclusion.}
In conclusion, we report the experimental realization of a flux-staggered rhombic chain by engineering synthetic flux in the momentum-space lattice of ultracold atoms. Through the flexible control of the Raman-coupled lattice in the synthetic dimensions, we are able to tune
synthetic flux within each plaquette on-demand and in a time-dependent fashion.
In the time-independent regime, we have observed the localized breathing modes and biased oscillation.
Based on the flux-dependent local dynamics under the flat-band condition,
we demonstrate topological chiral transport through the Floquet engineering. For future studies, it would be interesting to extend the BFL configuration to more complicated structures, including the 2D topological networks~\cite{2DML} and more ingenious lattice configurations for manipulating localized states~\cite{PhysRevB.102.054301}. The tuning of long-range interactions intrinsic to the momentum-lattice~\cite{Xie2020, An2021a, Xiao2021, PhysRevLett.129.103401} would also provide possibilities to study the interplay between flat-band localization and synthetic flux in the strongly correlated regime.

\begin{acknowledgments}
{\it Acknowledgement:}
We acknowledge the support from the National Key Research and Development Program of China under Grant No.2023YFA1406703 and No. 2022YFA1404203, the National Natural Science Foundation of China under Grants Nos. U21A20437, 12074337, 11974331, and 12374479, the Natural Science Foundation of Zhejiang Province under Grant No. LR21A040002, the Fundamental Research Funds for the Central Universities under Grant No. 2021FZZX001-02  and 226-2023-00131, the China Postdoctoral Science Foundation under Grant No. 2023M733122, and the Science Specialty Program of Sichuan University under Grant No. 2020SCUNL210.
\end{acknowledgments}

\bibliographystyle{apsrev4-1}
\bibliography{SABcage}

\end{document}